\documentclass[prb,twocolumn,superscriptaddress,citeautoscript,floatfix]{revtex4-1}
\usepackage{amssymb}
\usepackage{amsbsy}
\usepackage{amsmath}
\usepackage{bm}
\usepackage{color}
\usepackage{textcomp}
\usepackage{gensymb}
\usepackage{multirow}
\usepackage{array}
\usepackage{booktabs}
\usepackage{tabularx}
\usepackage{upgreek}
\usepackage{epsfig,psfrag,amsopn}
\usepackage{mathrsfs}
\usepackage{float}
\usepackage{graphicx}
\usepackage{ulem}
\usepackage{subcaption}
\usepackage[breaklinks,colorlinks = true,linkcolor = blue,urlcolor=cyan,citecolor=blue]{hyperref}
\begin{document}

\title{Strong coupling of lattice and orbital excitations in quantum magnet \texorpdfstring{Ca$_{10}$Cr$_7$O$_{28}$}{}: Anomalous temperature dependence of Raman phonons}

\author{Srishti Pal}
\email[E-mail:~]{srishtipal@iisc.ac.in}
\affiliation{Department of Physics, Indian Institute of Science, Bengaluru 560012, India}

\author{Arnab Seth}
\affiliation{International Centre for Theoretical Sciences, Tata Institute of Fundamental Research, Bengaluru 560089, India}
\affiliation{School of Physics, Georgia Institute of Technology, Atlanta, GA 30332, USA}

\author{Anzar Ali}
\affiliation{Indian Institute of Science Education and Research (IISER) Mohali, Knowledge City, Sector 81, Mohali 140306, India}

\author{Yogesh Singh}
\affiliation{Indian Institute of Science Education and Research (IISER) Mohali, Knowledge City, Sector 81, Mohali 140306, India}

\author{D. V. S. Muthu}
\affiliation{Department of Physics, Indian Institute of Science, Bengaluru 560012, India}

\author{Subhro Bhattacharjee}
\affiliation{International Centre for Theoretical Sciences, Tata Institute of Fundamental Research, Bengaluru 560089, India}

\author{A. K. Sood}
\affiliation{Department of Physics, Indian Institute of Science, Bengaluru 560012, India}

\date{\today}

\begin{abstract}

We report low-temperature Raman signatures of the Heisenberg quantum magnet Ca$_{10}$Cr$_7$O$_{28}$, showing clear anomalies in phonon mode frequencies and linewidths below $\sim$100 K. This crossover temperature lies in between the Jahn-Teller (JT) temperature scale ($>$ room temperature) and the temperature scale associated with the spin exchange interactions ($<$ 12 K). Our experimental observation is well captured by a novel secondary JT transition associated with a cooperative reorientation of the orbitals giving rise to anomalies in the temperature dependence of Raman frequencies and linewidths. Such orbital reorganisation, in turn, affects the spin-spin exchange interactions that decide the fate of the magnet at lower temperatures and hence provide important clues to understand the energetics of the possible lower temperature quantum paramagnetic phase.

\end{abstract}

\maketitle

Recent advances have invigorated vitality to the intricate interplay of different degrees of freedom giving rise to intervening coupling schemes in frustrated spin systems~\cite{Ramirez1994,Balents2010,Liu2016}. In particular, the role of these couplings in shaping the fate of geometrically frustrated spin-1/2 antiferromagnets -- originally conceptualized~\cite{Anderson1973} as suitable platforms for realising quantum spin liquid (QSL) states~\cite{Savary2016,Lee2008,Zhou2017,Knolle2019} -- have been thoroughly investigated both experimentally as well as on theoretical grounds on a large number of  known {\it geometrically frustrated motifs} such as triangular~\cite{Shimizu2003,Yamashita2009,Itou2007,Li2015}, kagom\'{e}~\cite{Helton2007,Fak2012,Fu2015}, or pyrochlore~\cite{Gingras2014,Scheie2020}. One important factor bearing the potential to alter the spin exchange interactions in these frustrated magnets with complex structures is orbital interactions~\cite{Koo2000}. Some of the magnetic perovskites like KCuF$_3$, LaMnO$_3$, and BaVS$_3$ had previously been studied for their coupling between the magnetic and orbital orderings responsible in minimizing the destabilizing two-electron two-orbital interactions~\cite{Oles2000,Whangbo2002,Whangbo2002-2}.

In this regard, the quasi-two-dimensional Heisenberg quantum magnet Ca$_{10}$Cr$_7$O$_{28}$~\cite{Balz2016,Balz2017,Balz2017-2,Alshalawi2022} on bi-layer kagome lattice presents a rather unique case disparate from the previously studied candidate geometrically frustrated QSLs. The compound with actual stoichiometry Ca$_{10}$(Cr$^V$O$_4$)$_6$(Cr$^{VI}$O$_4$)~\cite{Arcon1998} contains magnetically isolated distorted kagome bilayers of spin-$\frac{1}{2}$ Cr$^{5+}$ ions with both ferromagnetic (FM) and antiferromagnetic (AFM) isotropic Heisenberg exchange couplings~\cite{Balz2017,Balodhi2017}. The magnetic Hamiltonian constructed as a combination of experimental phenomenology and first principle calculations that accounts for the inequivalent FM (significantly stronger) and AFM isotropic Heisenberg couplings ($\Sigma J \lesssim$ 1 meV)~\cite{Balz2017-2} results in a mean-field Curie-Weiss temperature of $\theta_{CW} \approx$ 4 K~\cite{Balodhi2017} that is corroborated by experiments which show a broad cusp like feature around $T \approx$ 3.1 K in the magnetic specific heat~\cite{Balz2016}. However, the system shows no sign of long-range magnetic order or any spin-glass freezing down to 19 mK with a fluctuating spin liquid ground state as confirmed from bulk susceptibility, heat capacity, $\mu$SR, or neutron scattering measurements~\cite{Balz2016,Balz2017,Balz2017-2,Ni2018,Sonnenschein2019} as well as from theoretical studies~\cite{Damle2018,Sonnenschein2019,Pohle2021}. Due to the complex structure of this compound, more careful observation needs to be performed to disentangle the contribution of spin and orbital moments to the underlying magnetic Hamiltonian, and also possibly the role of lattice degrees of freedom. 

In this letter, we report our Raman scattering results on the Ca$_{10}$Cr$_7$O$_{28}$ system down to $\sim$4 K, revealing strong anomalies in temperature dependence of phonon frequencies and linewidths, at the crossover temperature $T_C \sim$100 K, much above the temperature scale ($\sim$10 K) associated with the spin-exchange interactions-- hence cannot arise from the non-trivial spin-phonon coupling since the spins, for all practical purposes are deep inside the thermal paramagnet,{\it i.e.}, effectively at infinite temperature ($T/\Theta_{CW}\gg 1)$. Therefore, these Raman anomalies are very much different from the temperature-induced magnetic ordering transitions~\cite{Dediu2000,Zhang2001} which are strictly absent in the Ca$_{10}$Cr$_7$O$_{28}$ system~\cite{Balz2017}. This raises the central question about the origin of this energy-scale and the associated Raman-active phonon renormalisation and here we show that this is naturally attributed to phonon-orbital coupling via a cascade of Jahn-Teller effects-- both single ion and cooperative. 

To begin the search for the physics of $100$ K energy scale, it is worth noting that the system contains Jahn-Teller (J-T) active~\cite{JT1937} Cr$^{5+}$ ion sites offering moderately distorted Cr$^V$O$_4$ tetrahedra even at room temperature and down to 2 K without any further structural distortion throughout the temperature range as reported in earlier studies~\cite{Gyepesova2013, Balz2017}. Therefore, the 100 K phonon anomalies cannot be attributed to the thermal order-disorder crossover from static to dynamic J-T effect which is associated with symmetry raising type structural transitions~\cite{Carron2001,Carron2001-2,Malcherek2017}. Hence, we turn our focus on the co-operative J-T effect driven by the interaction between localized orbitals and the crystal lattice~\cite{Gehring1975}. Such Raman fingerprints in terms of splitting of Raman bands for the co-operative J-T effect have previously been reported in rare-earth compounds like DyVO$_4$, DyAsO$_4$, and TbVO$_4$~\cite{Elliott1972} where the transitions are associated with only small lattice strains and splitting of the ground electronic states of the active ions. Here we present our detailed experimental observation of this intermediate temperature phenomena and support it with theoretical calculations under the framework of co-operative J-T effect to derive the phonon renormalization induced by an orbital reordering phenomenon~\cite{Ament2011,Saitoh2001} emerging from significant coupling between the vibrational and orbital degrees of freedom. Such reordering has direct implication on the spin physics of this frustrated magnet at even lower temperature via drastic renormalisation of the spin-spin exchanges.

\begin{figure}[htp!]
\centering
\includegraphics[width=80mm,clip]{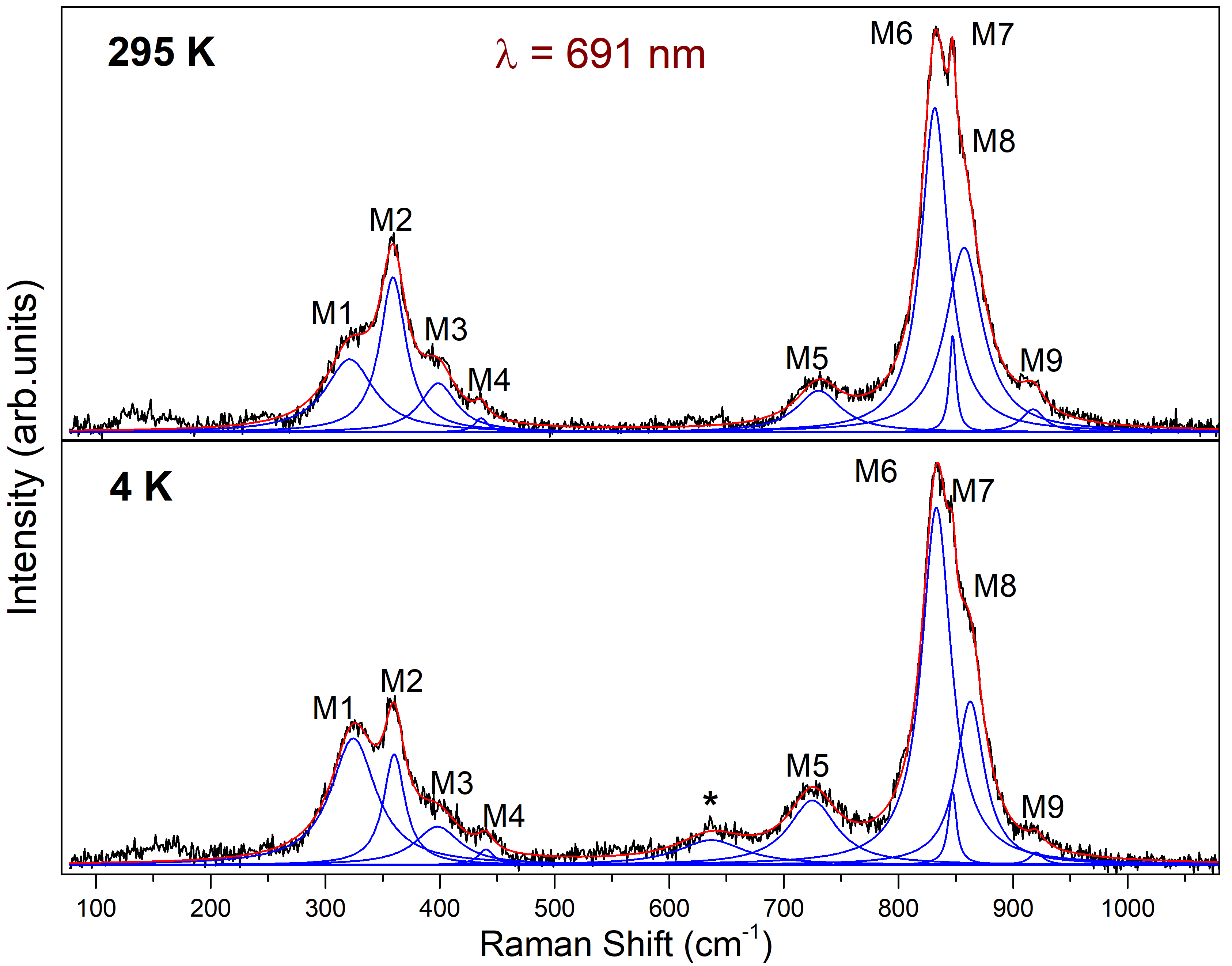}
\caption{\small The two panels show phonon fits to the Raman spectra at 295 K and 4 K. Experimental data are indicated by black curves. Blue and red curves are individual phonon modes and the cumulative fits, respectively. `*' represents the new mode appearing at 4 K.}
\label{figure1}   
\end{figure}

\begin{figure}
\centering
\includegraphics[width=80mm,clip]{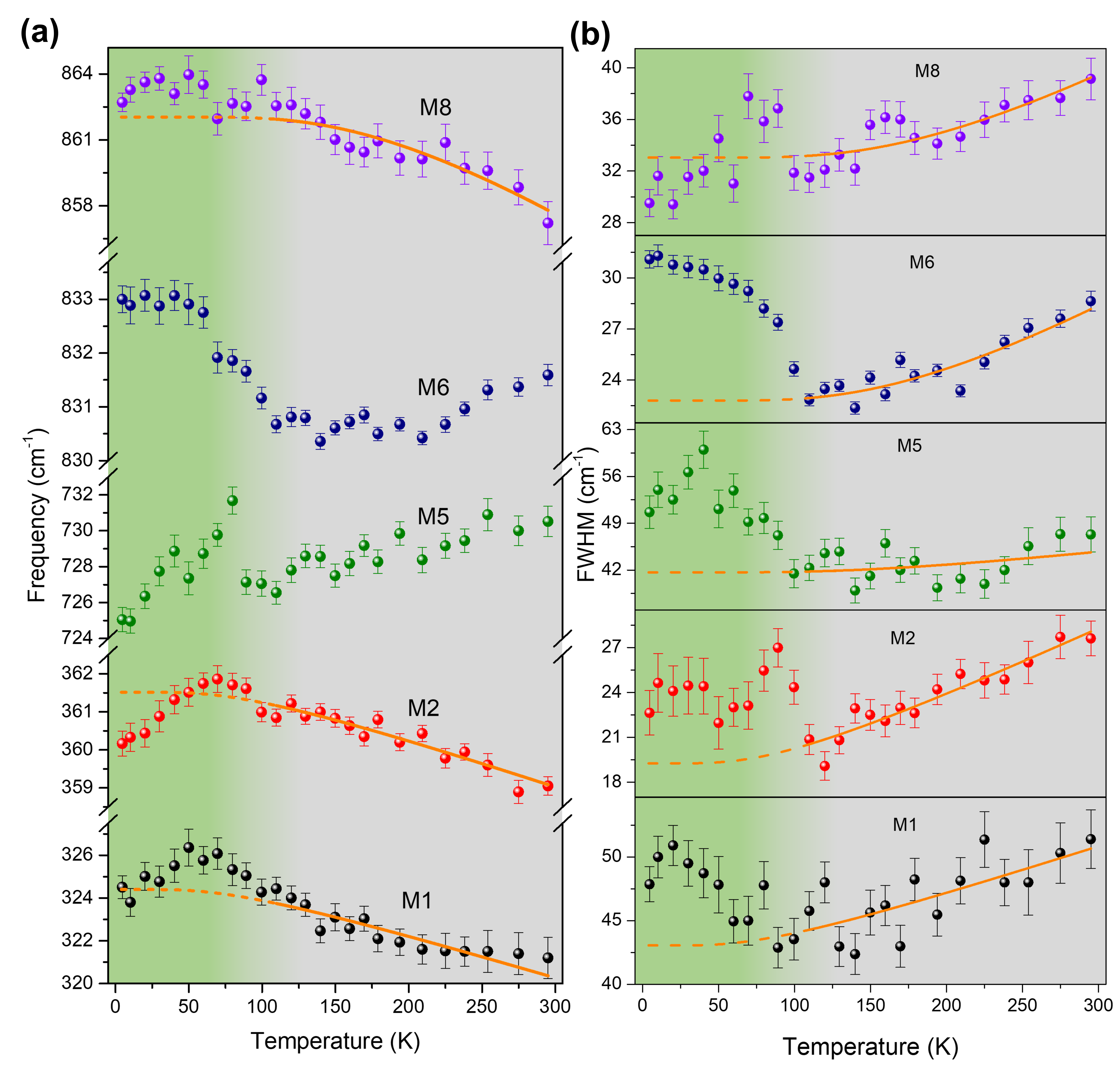}
\caption{\small {Temperature dependence of \textbf{(a)} frequencies and \textbf{(b)} linewidths of selected phonon modes. The orange curves are fits to cubic anharmonic model with $\omega_{anh}^{(p)}=\omega_0+A\left[1+2n\left(\frac{\omega_0}{2}\right)\right]$ and $\Pi_{anh}^{(p)}=\Gamma_0+B\left[1+2n\left(\frac{\omega_0}{2}\right)\right]$. Details of the parameters are given in the SI~\cite{Suppl}. }}
\label{figure2}   
\end{figure}

Raman spectra of Ca$_{10}$Cr$_7$O$_{28}$ at 4 K and 295 K are shown in Fig.~\ref{figure1} along with the phonon fits. Factor group analysis of trigonal (\textit{R3c}) Ca$_{10}$Cr$_7$O$_{28}$ yields 139 Raman active phonon modes ($\Gamma_{Raman}$ = 46 $A_{1g}$ + 93 $E_g$) among which 9 modes could be detected at 295 K in the frequency range 75 - 1000 cm$^{-1}$. From earlier reports on compounds containing Cr$^V$O$_4$ tetrahedra~\cite{Aoki2000,Aoki2001}, we assign the low (M1-M4) and high (M5-M9) frequency vibrational bands of Ca$_{10}$Cr$_7$O$_{28}$ as bending and stretching modes of the Cr$^V$O$_4$ tetrahedra, respectively.

The phonon modes are fitted with symmetric Lorentzian profile function for the entire range of temperature. The phonon spectrum remains unchanged with decreasing temperature except only at the lowest temperature of 4 K where one new weak mode at $\sim$640 cm$^{-1}$ [indicated by (*) in the lower panel of Fig.~\ref{figure1}] could be detected. Appearance of this weak mode may be associated with crossovers in the spin channels~\cite{Damle2018,Pohle2021} whose fingerprints could also be observed in earlier specific heat studies~\cite{Balz2016}.

The temperature evolution of frequencies and FWHMs for selected phonon modes are shown in Fig.~\ref{figure2}(a)-(b). The solid orange curves are fits from 100 K to 295 K to the simple cubic anharmonic model~\cite{Klemens1966} (see SI~\cite{Suppl} for fitting details). The dashed ones are their extensions to lower temperature ranges. It is worth noting that except for M8, frequencies and FWHMs of all other modes exhibit anomalous behaviour with temperature. While frequencies of M1, M2 and FWHMs of M1, M2, M5, and M6 show clear deviations from the expected cubic anharmonicity below $\sim$100 K, frequencies of M5 and M6 modes are anomalous throughout the entire temperature range and hence, could not be fitted with the cubic anharmonic model. While M5 frequency shows a discontinuity around 100 K with a change in the slope, M6 frequency experiences slope change around 100 K with striking similarity in its temperature profile with that of its linewidth. Apart from the strong anomalies in frequencies and linewidths, the phonon modes show subtle anomalous behaviour in the temperature dependence of their integrated susceptibilities as shown in SI~\cite{Suppl}.

As remarked above, the crossover temperature of $\sim$100 K associated with the strong phonon anomalies in the Ca$_{10}$Cr$_7$O$_{28}$ system is too high to be linked with the spin-exchange interactions with a temperature scale of $\sim$10 K and below. Also, the neutron time-of-flight (TOF) powder diffraction patterns~\cite{Balz2017} confirmed absence of any structural phase transition down to 2 K. However, an interesting observation regarding Ca$_{10}$Cr$_7$O$_{28}$ crystal system is that its Cr$^V$O$_4$ tetrahedra are distorted at room temperature, bearing four different Cr-O bond lengths (see SI~\cite{Suppl}). This can be understood by noting that the Cr$^{5+}$, in a tetrahedral crystal field has one electron in the $e_g$ orbitals and hence is a Jahn-Teller (J-T) active ion~\cite{JT1937}. The degeneracy of the atomic orbitals are, therefore, susceptible to J-T splitting associated with the distortion of the tetrahedra (see Figure~\textcolor{blue}{S2} in SI~\cite{Suppl}). 

The above experimental phenomenology pose the following question: Is the $\sim$100 K scale seen by the phonons related to the J-T rearrangement of the orbitals due to various orbital-lattice couplings? Indeed, the different vibrational modes allow for such {\it secondary cooperative} J-T distortions. Below, we explore the possibility of such physics to explain the phenomenology of Ca$_{10}$Cr$_7$O$_{28}$ via minimal symmetry allowed Hamiltonians.

A minimal model that captures the above J-T distortion includes the spin-1/2, $S^\alpha_i$, and the two $e_g$-orbitals ($(|d_{3z^2-r^2}\rangle,|d_{x^2-y^2}\rangle)\equiv (|+\rangle,|-\rangle)$) $\tau^\alpha_i$ on the Cr$^{5+}$ sites as well as the Raman active phonons, $\epsilon^{p}_i$ ($p$ denotes the normal modes) with appropriate symmetries. The associated spin-orbital-phonon Hamiltonian is given by $H=H_{S,\tau,\epsilon}+H_{\epsilon}$, where $H_{\epsilon}$ is the harmonic phonon Hamiltonian and 
\begin{align}
    H_{S,\tau,\epsilon}=\sum_{\langle ij\rangle}\left[K^{\alpha\beta}_{ij}+J^{\alpha\beta}_{ij} {\bf S}_i\cdot {\bf S}_j\right] \tau^\alpha_i\tau^\beta_j+\sum_i\Gamma^{p,\beta}\epsilon_i^{p} \tau_i^\beta
    \label{eq_sopham}
\end{align}
with $J^{\alpha\beta}_{ij}$ and $K_{ij}^{\alpha\beta}$ denote the bare spin-spin and orbital-orbital exchange interactions respectively between two neighbouring Cr$^{5+}$ ions;  $\Gamma^{p,\beta}$ is the coupling of the orbitals with the phonon modes. Note that, due to the absence of atomic spin-orbit coupling (expected to be small for 3d transition metals), the spin-spin interactions are rotationally symmetric while the dependence on the orbital labels is constrained by lattice symmetries. The second term denotes the symmetry allowed linear coupling between the orbitals and the phonon modes that is present in a J-T active ion and is responsible for the distortion via $\epsilon^{p}=-\frac{\Gamma^{{p},\beta}}{C_{p}}\tau^\beta$ ($C_p$ is the spring constant of the phonon) along the direction determined by largest J-T coupling $\Gamma^{p,\beta}$ for the softest elastic mode. This in-turn decides the specific splitting of the orbitals, {\it i.e.}, $\langle\tau^\alpha\rangle\neq0$.

Below the temperature scale of this {\it primary J-T} splitting, the other J-T coupling constants, as well as the orbital-coupling scale can induce further {\it secondary J-T} transitions which can be rendered cooperative due to $K_{ij}^{\alpha\beta}$. These successive J-T transitions, therefore, lead to rearrangement of the orbital ordering which we, as explained below, attribute to the experimentally observed  Raman anomalies around $T\sim 100$ K. 

\paragraph*{Orbital fluctuations at intermediate temperatures.} Consider the intermediate temperature range where the Raman anomalies are observed : $\Theta_{CW}(\sim 10~{\rm K})\ll T< T_H(>300~K)$, where $\Theta_{CW}\sim 10$ K and $T_H$ are respectively the Curie-Weiss  and the primary J-T transition temperatures. In this temperature range, the spins form a thermal paramagnet {\it i.e.}, $\langle{\bf S}_i\rangle=0$, and are completely incoherent. Therefore, their only effect is to renormalise the couplings $K_{ij}^{\alpha\beta}$ in Eq. \ref{eq_sopham} via short-ranged spin-spin correlations. However, the orbital ordering has set in due to the primary J-T effect. Let us assume that this ordering is along $\tau^z$ (we do not know the actual direction of distortion from the powder samples). Therefore, the effective Hamiltonian in the intermediate temperature range is given by
\begin{align}
    \tilde H=&\sum_{\langle ij\rangle}\mathcal{K}^{\alpha\beta}_{ij} \tau^\alpha_i\tau^\beta_j+\Delta_1\sum_{i}\tau^z_i+\sum_i\Gamma^{p,\beta}\epsilon_i^{p}\tau^\beta_i+\tilde{H}_{\epsilon}
    \label{eq_horb}
\end{align}
where $\mathcal{K}_{ij}^{\alpha\beta}=K_{ij}^{\alpha\beta}+J_{ij}^{\alpha\beta}\langle{\bf S}_i\cdot {\bf S}_j\rangle$ is the effective exchange for the orbitals; $\Delta_1$ characterises the splitting of the orbitals due to the primary J-T effect. In principle, there will also be a symmetry allowed term of the form $\Delta_2\sum_i\tau^x_i$ coming from the lattice distortion at the primary J-T transition, but the effect of such terms is straight forward-- they smear out sharp features of phase transitions arising from Eq.~\ref{eq_horb} and makes way for smooth crossover as seen in experiments~\cite{Carron2001,Zhang2001,Malcherek2017}.

In Eq. \ref{eq_horb}, the sum $p$ now runs over the other phonon modes that can potentially lead to secondary co-operative J-T mediated reordering via  $\mathcal{K}_{ij}^{\alpha\beta}$; and $\tilde{H}_{\epsilon}$ is the harmonic phonon Hamiltonian for the phonons in the distorted state below the primary J-T transition. Assuming a single secondary mode favouring an ordering along $\tau^x$, the mean-field phase diagram~\cite{Stinchcombe_1973} is easy to work out and given in the SI~\cite{Suppl} (see Figure~\textcolor{blue}{S3}). In the mean-field approximation, the orbital reordering sets in below a temperature, $T_c=\left(\mu^{xx}-\tilde{\mathcal{K}}^{xx}\right)\frac{\delta}{\tanh^{-1}(\delta)}$, where $\delta=\Delta_1/\left(\mu^{xx}-\tilde{\mathcal{K}}^{xx}\right)$, $\tilde{\mathcal{K}}^{xx}=\mathcal{K}^{xx}D$ ($D$ = coordination number) and $\mu^{xx}=\frac{\Gamma^{p,x}\Gamma^{p,x}}{C_{p}}$ are effective coupling constants. We associate $T_c$ to $\sim 100$ K with the experimentally observed onset scale of the phonon anomalies.

\begin{figure}
    \centering 
    \includegraphics[width=8.2cm]{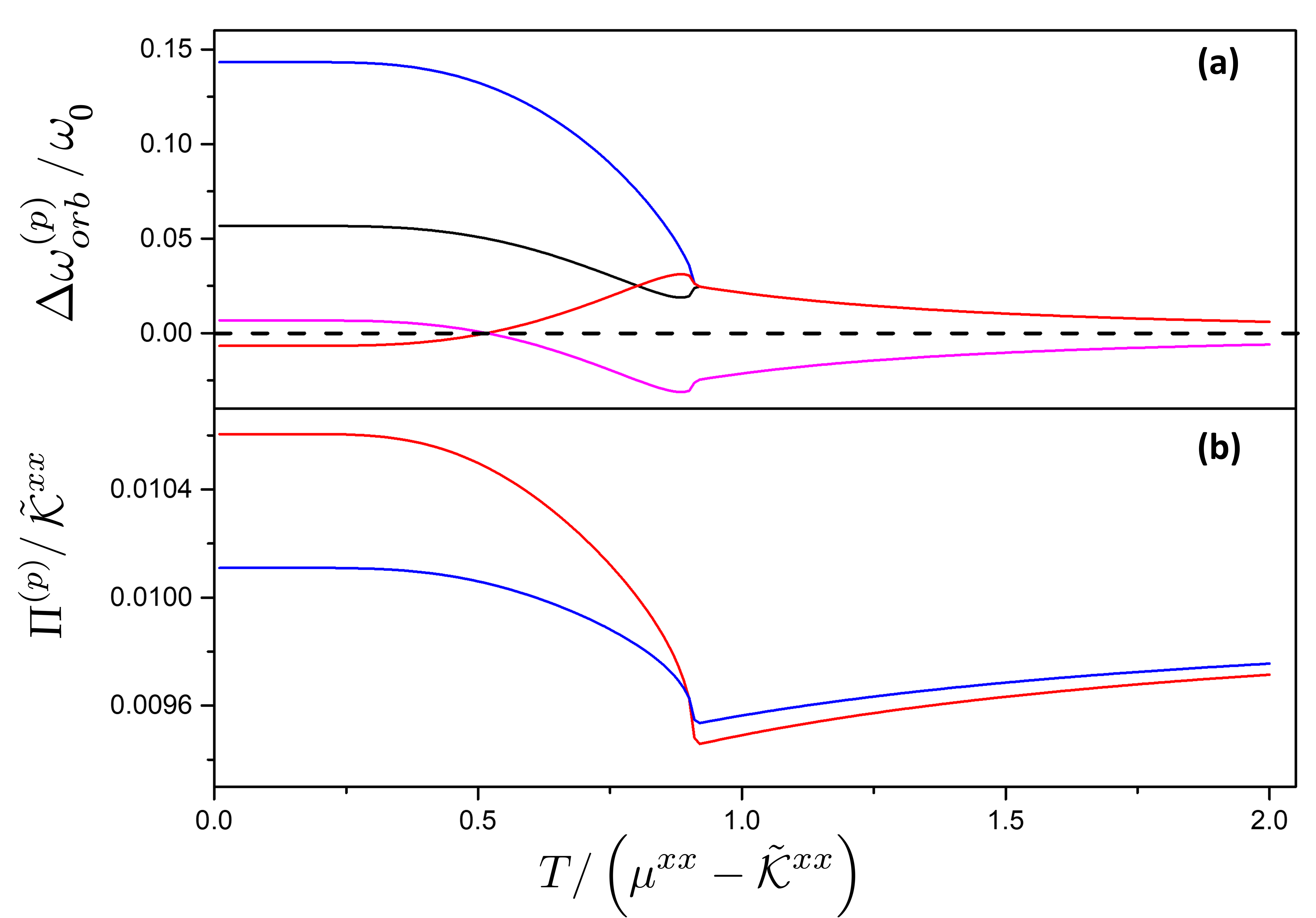}
\caption{{\bf (a)} Frequency shift due to the orbital reordering. Blue, black, red, and magenta curves are obtained by choosing $(B^{xx;pp},B^{zz;pp},B^{xz;pp})\omega_0$ to be $(0.1,0.1,0.1)$, $(0.1,0.1,-0.1)$, $(-0.1,0.1,-0.1)$, and $(0.1,-0.1,0.1)$, respectively. {\bf (b)} Linewidth renormalisation of phonon. For blue and red curves, we choose $\Gamma_0=0.01\tilde{\mathcal{K}}^{xx},~ \mathcal{M}_{xx}=\mathcal{M}_{xz}=-0.7\tilde{\mathcal{K}}^{xx}$ and $\Gamma_0=0.01\tilde{\mathcal{K}}^{xx},~ \mathcal{M}_{xx}=-0.6\tilde{\mathcal{K}}^{xx},~\mathcal{M}_{xz}=-0.4\tilde{\mathcal{K}}^{xx}$, respectively and set all other parameters to zero. For both the panels, $\delta=0.5$.}
\label{fig_theory}
\end{figure}

\paragraph*{Phonon anomaly due to secondary J-T transition.} In the intermediate temperature regime, the phonon anomalies observed in our vibrational Raman scattering experiments can be analysed using the phonon-orbital coupling Hamiltonian which is given in details in the SI (see Eq.~\textcolor{blue}{S5}). For this, we note that in addition to $\Gamma^{p,\alpha}$, the orbital-phonon coupling has another source {\it i.e.}, the coupling constant, $\mathcal{K}_{ij}^{\alpha\beta}$, which are dependent on the dynamic distortions of the tetrahedra and hence we have, similar to the usual magnetoelastic coupling,
\begin{align}
    \mathcal{K}^{\alpha\beta}_{ij}=\bar{\mathcal{K}}_{ij}^{\alpha\beta}+A_{ij}^{\alpha\beta;p}(\epsilon_i^{p}+\epsilon_j^{p}) +B_{ij}^{\alpha\beta;pq}\epsilon_i^{p}\epsilon_j^{q}
    \label{eq_expansion}
\end{align}
where $A_{ij}^{\alpha\beta;{p}}$ and $B_{ij}^{\alpha\beta;pq}$ are coupling constants whose different components are constrained by the residual symmetries. Using this in the orbital Hamiltonian, $H_{orb}$, we obtain the orbital-phonon coupling that is central to the vibrational Raman scattering (see SI~\cite{Suppl}). The phonon renormalistion can then be computed perturbatively due to these interactions.

The renormalisation of phonon frequency and linewidth are respectively given by,
\begin{align}
    &\Delta\omega^{(p)}=\omega^{(p)}-\omega_0=\Delta\omega^{(p)}_{anh}+\Delta\omega^{(p)}_{orb}\\
    &\Pi^{(p)}=\mid\Pi^{(p)}_{anh}+\Pi^{(p)}_{orb}\mid
\end{align}
where the subscript ``${anh}$" and ``${orb}$" represent the contributions due to the anharmonic effects and orbital reordering, respectively. While the anharmonic contribution to the phonon parameters are usually determined from the fitting of the experimental data at high temperatures (see SI~\cite{Suppl}), the orbital contribution can be computed within Einstein (independent bond) approximation. The leading order renormalisation of the frequency is given by
\begin{align}
    \Delta\omega_{orb}^{(p)}&\propto B_{ij}^{\beta\gamma;pp}\langle\tau_i^\beta\tau_j^\gamma\rangle
\end{align}
where $\langle\cdots\rangle$ denotes the averaging over a Gibbs ensemble at temperature $1/\beta$. The phonon frequency renormalization due to orbital reordering, therefore, is determined by the equal-time orbital-correlators on nearest neighbours. Within our minimal model, approximating $\langle\tau^x_i\tau^x_j\rangle\approx\langle\tau^x\rangle^2$, $\langle \tau^z_i\tau^z_j\rangle\approx \langle\tau^z\rangle^2$ and $\langle\tau^x_i\tau^z_j\rangle\approx\langle\tau^x\rangle\langle\tau^z\rangle$ below $T_c$, we obtain results that are plotted in Fig.~\ref{fig_theory} (top panel) as a function of temperature for various representative choices of the coupling constants. In particular, as is clear from these plots that the softening or hardening of the phonon is determined by the sign of the coupling $B_{ij}^{\beta\gamma;pp}$. We see both these behaviour for different phonons that are consistent with the experimental observation [see Fig.~\ref{figure2}(a)]. An estimate of these coupling constants, however, requires a more microscopic calculation which is beyond the purview of the present symmetry-based arguments.

Turning to the linewidths, we calculate it via well-known methods of diagrammatic perturbation theory whence the linewidth is given by the imaginary part of the phonon self-energy arising due to its scattering with $\tau^\alpha_i$s and determined by, to the leading order, the dynamic orbital correlation functions $\mathcal{C}_{\mu\nu}({\bf r}, \tau)=\langle\hat{T}\left(\tau_{\bf r}^\mu(\tau)\tau_{\bf 0}^\nu(0)\right)\rangle-\langle\tau^\mu_{\bf r}(\tau)\rangle\langle\tau^\nu_{\bf 0}(0)\rangle$.

The resultant leading order Raman linewidths are given by~\cite{arnab_thesis}
\begin{align}
    \Pi_{orb}^{(p)}\left(\omega\right)\sim\sum_{\mu,\nu=x,z}\mathcal{M}_{\mu\nu}\text{Im}\left[\tilde{\mathcal{C}}_{\mu\nu}\left({\bf k}=0,\omega+i0^+\right)\right]
    \label{eq_lw_final}
\end{align}
where $\tilde{\mathcal{C}}_{\mu\nu}({\bf k}, \omega+i0^+)$ is the fourier transform of $\mathcal{C}_{\mu\nu}({\bf r}, \tau)$ and $M_{\mu\nu}$ is the form factor whose form can be found in the SI~\cite{Suppl} (see Eq.~\textcolor{blue}{S8}). For simplicity, in Fig. \ref{fig_theory} (bottom panel) we plot the temperature dependence of the linewidth assuming these coefficients to be temperature independent, and therefore this is now completely controlled by the two-point dynamic correlators of the orbitals. We note that such behaviour is in direct conformity with the experimental observation [see Fig. \ref{figure2}(b)].

In conclusion, we have explored temperature-induced Raman anomalies in polycrystalline samples of the Heisenberg quantum magnet Ca$_{10}$Cr$_7$O$_{28}$. The phonon mode frequencies, linewidths, and integrated intensities reveal clear anomalies across $\sim$100 K, a temperature scale much higher than the one associated with the spin-exchange interactions of the system ($<$ 10 K). Considering the fact that the system is Jahn-Teller distorted above room temperature, we develop our theoretical understanding to realize the Raman anomalies to originate from an orbital reordering phenomenon, renormalizing the phonon self-energy via cooperative Jahn-Teller effect. As we have used polycrystalline sample in our studies, we could not do an extensive assignment of different phonon modes to different vibrations (symmetries). In future, polarization-dependent Raman studies on single crystals can help to understand the origin of different nature of anomalies for different phonon modes. Also, resonant inelastic X-ray scattering measurements can be performed on the system as a function of temperature to capture the orbital ordering. In the passing, we note that, once the orbitals are ordered at low temperature ($T< 100$ K), the effective spin-Hamiltonian is obtained from Eq. \ref{eq_sopham} as $H_{\rm spin}=\sum_{ij} \mathcal{J}_{ij} {\bf S}_i\cdot {\bf S}_j+\cdots$ where $\mathcal{J}_{ij}=\sum_{\alpha\beta}J_{ij}^{\alpha\beta}\langle\tau^\alpha_i\rangle\langle\tau^\beta_j\rangle$ are the effective low-temperature spin-spin exchanges and $\cdots$ refer to other constant terms. This should result in variation of the Curie-Weiss temperature and can account for the low effective spin-spin exchanges observed in Ca$_{10}$Cr$_7$O$_{28}$ which, in turn, dictates the fate of the spin state at lower temperatures. Thus, the secondary J-T physics plays an important role in ultimately deciding the the possible QSL state in the material.

\begin{acknowledgments}

{AKS thanks DST for financial support under the National Science Chair Professorship. AS and SB acknowledge funding from Max Planck Partner group Grant at ICTS, Swarna jayanti fellowship grant of SERB-DST (India) Grant No. SB/SJF/2021-22/12 and the Department of Atomic Energy, Government of India, under Project No. RTI4001.}

\end{acknowledgments}

\end{document}